# A Spatio-Temporal Feature Fusion EEG Virtual Channel Signal Generation Network and Its Application in Anxiety Assessment

Shangqing Yuan, Wenshuang Zhai, and Shengwen Guo

*Abstract*—To address the issue of limited channels and insufficient information collection in portable EEG devices, this study explores an EEG virtual channel signal generation network using a novel spatio-temporal feature fusion strategy. Based on the EEG signals from four frontal lobe channels, the network aims to generate virtual channel EEG signals for other 13 important brain regions. The architecture of the network is a two-dimensional convolutional neural network and it includes a parallel module for temporal and spatial domain feature extraction, followed by a feature fusion module. The public PRED+CT database, which includes multi-channel EEG signals from 119 subjects, was selected to verify the constructed network. The results showed that the average correlation coefficient between the generated virtual channel EEG signals and the original real signals was $0.6724 \pm 0.02$, with an average absolute error of $3.9470 \pm 1.23$. Furthermore, the 13 virtual channel EEG signals were combined with the original EEG signals of four brain regions and then used for anxiety classification with a support vector machine. The results indicate that the virtual EEG signals generated by the constructed network not only have a high degree of consistency with the real channel EEG signals but also significantly enhance the performance of machine learning algorithms for anxiety classification. This study effectively alleviates the problem of insufficient information acquisition by portable EEG devices with few channels.

## I. Introduction

ELECTROENCEPHALOGRAM (EEG) captures millisecond-scale neuronal dynamics and is widely employed in both neuroscience research and clinical neuropsychiatry. Multi-channel EEG devices record the electrical activity of neurons in various brain regions through multiple electrodes placed on the scalp, providing information on brain activity and function. These devices have been used for anxiety state assessment or anxiety disorder identification [1]-[5], attention deficit hyperactivity disorder (ADHD) [6]-[8], and depression [9], [10] , etc.

Clinical-grade high-density systems offer comprehensive spatial coverage and high signal fidelity yet are cumbersome, expensive, and impractical for ambulatory or home-based monitoring. However, their numerous electrodes, large size, inconvenience in wearing, strong discomfort, and high cost limit their application in psychological assessment and treatment, clinical diagnosis, and home health monitoring. To address these issues, the portable EEG devices with fewer electrodes have emerged as a cost-effective and convenient alternative. For instance, Al-Kaf et al. [11] used the Mindwave Mobile 2 portable EEG device to collect single-channel EEG signals, demonstrating that the device could provide feedback on the brain's relaxation level, helping users alleviate stress and anxiety in a home environment. Arsalan et al. [12] used the Interaxon Muse to obtain four-channel EEG signals from the forehead, calculated temporal features, and applied a random forest classifier for anxiety classification with an accuracy of 78.5%. Arsalan et al. [13] also used t-tests and analysis of variance to select features from channels with significant differences for anxiety binary and ternary classification, achieving accuracy rates higher than 80% and verifying the effectiveness of portable devices in anxiety classification tasks. Additionally, Cannard et al. [14] compared the Interaxon Muse head-worn device with an EEG system (64-channel Biosemi Active Two) to validate the reliability of the Interaxon Muse device in anxiety recognition. Sidelinge et al. [15] applied the Muse 2 to collect four-channel EEG signals in a home environment and compared them with professional EEG acquisition devices (HD-EEG), demonstrating that the device could stably record EEG feature changes related to anxiety levels in the Alpha frequency.

These studies show that portable EEG devices with fewer channels have attracted widespread attention due to their convenience and low cost advantages, demonstrating unique application value and potential. Although these devices effectively overcome the shortcomings of multi-channel EEG technology, their limited signal coverage of brain regions leads to insufficient information, affecting the accuracy of psychological research and clinical assessment. Therefore, addressing the insufficiency of information in portable EEG

This work was supported by Natural Science Foundation of Guangdong Province, China (No. 2023A1515011607, 2025A1515011746), Guangzhou Key Research and Development Program China (No. 2023B03J1335) (Corresponding author: Shengwen Guo).

S. Yuan and W. Zai are with the School of Automation Science and Engineering, South China University of Technology, Guangzhou, P.R. China, 510640. S Guo is with the School of Automation Science and Engineering, South China University of Technology, Guangzhou, P.R. China, 510640 (e-mail: shwguo@scut.edu.cn).

devices is crucial for ensuring their effective application.

In recent years, interpolation methods of EEG virtual channel signal are applied to enhance the spatial resolution of low-channel EEG devices. The interpolation methods mainly rely on the physical positional relationships between electrodes [16], [17]. For example, Courellis et al. [18] proposed a distance-based interpolation method that reconstructs missing EEG signals using metrics such as Euclidean distance, great-circle distance, and Elliptic Geodesic Length (EGL) through inverse distance weighting. Petrichella et al. [19] applied inverse distance weighting and spherical spline methods to address artifacts and missing signals in EEG data, finding that the spherical spline method has a clear advantage in terms of global reconstruction error. However, these interpolation methods consider only the physical locations of electrodes, assume spatial continuity and smoothness of the signals, which limits their ability to capture the complex, nonlinear, and dynamic nature of EEG. Their accuracy also heavily depends on the density and arrangement of electrodes.

The rise of deep learning has inaugurated a novel paradigm for the generation of virtual-channel EEG signals. For example, Corley et al. [20] proposed a deep EEG super-resolution (SR) method based on Generative Adversarial Networks (GAN), generating high spatial resolution signals from low-resolution EEG signals to reduce the demand for high-cost EEG devices. Kwon et al. [21] combined convolutional neural networks (CNN) with SR technology to improve the spatial resolution of EEG signals, exploring the dynamic brain activity with fewer electrodes. Svantesson et al. [22] proposed a CNN-based generation network that can recover or upsample signals from a few EEG channels, with generated signals outperforming traditional interpolation methods in terms of error, correlation, and visual assessment. Li et al. [23] proposed an improved GAN, Wasserstein Generative Adversarial Networks (WGAN), achieving higher training stability and improving the performance of motor imagery classification by learning key features from other channels, enhancing the user experience of brain-computer interfaces. Sun et al. [24] proposed the AMACW algorithm, which uses attention mechanisms to allocate channel weights. After evaluating EEG data with 1-5 bad channels and comparing it with the EEGLab interpolation method, the model was shown to effectively reconstruct bad channels. Chen et al. [25] designed a GAN architecture incorporating multi-head attention mechanisms, generating virtual EEG signals from a small amount of EEG data and validating the model's effectiveness through quantitative assessment, feature similarity, and classification tasks. Yang et al. [26] proposed an attention mechanism-based U-Net architecture, generating 28 virtual channel signals from 32 real EEG channel signals, with the generated signals showing excellent similarity in time-series graphs.

Although these studies demonstrate the significant advantages of deep learning methods in virtual EEG channel generation, most constructed networks focus on effectively extracting and learning the temporal features of each channel and the feature associations between each source channel signal and the generated virtual channel signal. They generally consider each source channel signal independently as network input. In reality, channels are not isolated from each other in both physical space and function. If the intrinsic correlations between source channels in time and space are ignored, the generated virtual channel signals will have significant deviations from the actual channel signals, resulting in low consistency and affecting subsequent analysis.

To address the issue of insufficient information collection by portable EEG devices due to their limited channels, We present STFPF-EEG GNet, a parallel spatio-temporal convolutional framework that (i) explicitly models intra-channel temporal dynamics and inter-channel spatial coherence, (ii) enforces consistency in both time and frequency domains via a composite loss, and (iii) demonstrates consistent performance gains on independent wearable-EEG data.

Additionally, to evaluate and validate the effectiveness and application value of the generated signals, machine learning algorithms are applied for anxiety classification on publicly available datasets to assess the performance of the classification algorithms.

## II. METHODS

### A. Network Architecture

The architecture comprises four stages: (1) input reshaping, (2) parallel temporal and spatial encoders, (3) bidirectional-LSTM temporal refinement, and (4) feature-fusion decoding. The temporal feature extraction module extracts temporal features for each channel signal in the time dimension, while the spatial domain feature extraction module performs convolution in the channel direction to capture the correlation features between channels. These two types of feature modules can be connected in either cascading or parallel way. In the cascading connection method, temporal features of each channel are first extracted, and then these features are input into the spatial domain feature extraction module for convolution operations to extract spatial domain features, with feature fusion achieved through concatenation. In the parallel connection method, two independent branches are designed, one for the temporal feature extraction module and the other for the spatial domain feature extraction module. Subsequently, the two types of features are fused through a feature fusion module.

This network architecture can utilize both the temporal features of each channel's EEG signal and the correlation features between channels, achieving the fusion of temporal and spatial domain features of EEG signals and thereby enhancing the network's learning and expression capabilities. Moreover, to capture the long-range forward and backward complex dependencies of signals in the time dimension, a bidirectional long short-term memory module (BiLSTM) is introduced into the network. Fig. 1 shows the parallel network structure, which is called the Spatio-temporal feature parallel fusion EEG generative network (STFPF-EEG GNet).

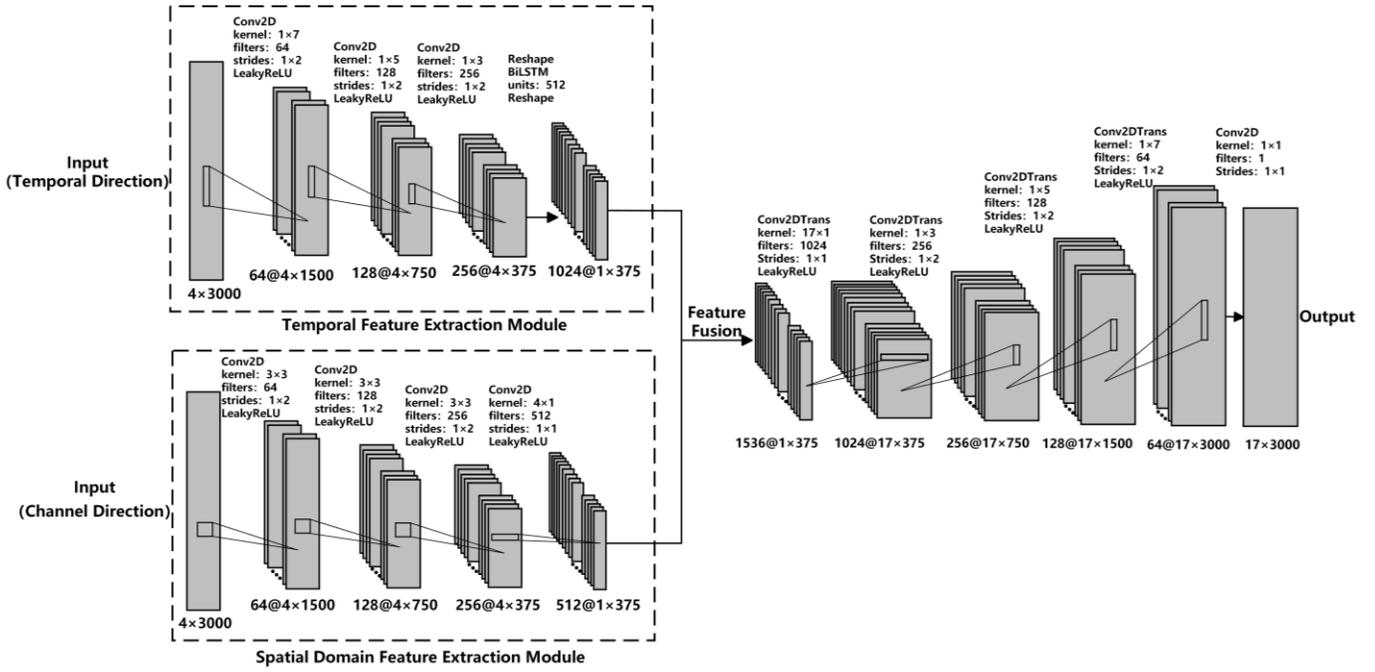

Fig. 1. Network Architecture.
Conv2D: 2D convolution, LeakyReLU: activation function, Reshape: dimension reshaping, Conv2DTrans: 2D transposed convolution, BiLSTM: bidirectional LSTM

TABLE I
NETWORK STRUCTURE PARAMETERS

| STFPF-EEG GNet | Layers | Filter numbers | Filter size | Strides | Padding | Activation |
|---|---|---|---|---|---|---|
| Temporal feature extraction module | Conv2D | 64 | (1, 7) | (1,2) | Same | LeakyReLU |
| | Conv2D | 128 | (1, 5) | (1,2) | Same | LeakyReLU |
| | Conv2D | 256 | (1, 3) | (1,2) | Same | LeakyReLU |
| | BiLSTM | 512 | — | — | — | — |
| Spatial domain feature extraction module | Conv2D | 64 | (3, 7) | (1,2) | Same | LeakyReLU |
| | Conv2D | 128 | (3, 5) | (1,2) | Same | LeakyReLU |
| | Conv2D | 256 | (3, 3) | (1,2) | Same | LeakyReLU |
| | Conv2D | 512 | (4, 1) | (1,1) | Vaild | LeakyReLU |
| Feature fusion and output module | Conv2DTrans | 1024 | (17,1) | (1,1) | Vaild | LeakyReLU |
| | Conv2DTrans | 256 | (1, 3) | (1,2) | Same | LeakyReLU |
| | Conv2DTrans | 128 | (1, 5) | (1,2) | Same | LeakyReLU |
| | Conv2DTrans | 64 | (1, 7) | (1,2) | Same | LeakyReLU |
| | Conv2D | 1 | (1, 1) | (1,1) | Vaild | — |

Conv2D: 2D convolution, Conv2DTrans: 2D transposed convolution, LeakyReLU: activation function, BiLSTM: bidirectional LSTM, Same: zero padding applied during convolution, Valid: no zero padding applied during convolution.

## B. Loss Function

To enhance the consistency between the generated signals and the original real signals, and guarantee fidelity in both domains, we minimize a weighted sum of time-domain MSE and frequency-domain MSE:

$$L = \alpha Loss_{TMSE} + \beta Loss_{FMSE} \quad (1)$$

where $Loss_{TMSE}$ and $Loss_{FMSE}$ are the mean squared errors in the time and frequency domains, respectively, and α and β are weight constants.

The mean squared error in the time domain is defined as:

$$Loss_{TMSE} = \frac{1}{N}\sum_{i=1}^{N}(f_i - g_i)^2 \quad (2)$$

where $f_i$ and $g_i$ amplitude of the original and generated signals, respectively, and N demotes their length.

The mean squared error in the frequency domain is defined as:

$$Loss_{FMSE} = \frac{1}{M}\sum_{k=1}^{M}(F_k - G_k)^2 \quad (3)$$

Where F and G represent the spectra of the original signal f and the generated signal g, respectively. k denotes the frequency, and M is the maximum frequency.

## C. Performance evaluation metrics

To evaluate the performance of the generative model, we use the mean absolute error (MAE) and the correlation coefficient (CC) to quantify the consistency between $f_i$ and $g_i$. The two

metrics are given as:

$$MAE = \frac{1}{N}\sum_{i=1}^{N}|f_i - g_i| \quad (4)$$

$$CC = \frac{\sum_{i=1}^{N}(f_i-\bar{f})(g_i-\bar{g})}{\sqrt{\sum_{i=1}^{N}(f_i-\bar{f})^2 \sum_{i=1}^{N}(g_i-\bar{g})^2}} \quad (5)$$

where $\bar{f}$ and $\bar{g}$ are the means of $f_i$ and $g_i$, respectively. A smaller MAE and higher correlation coefficient indicates closer similarity between the two signals.

### III. DATA AND EXPERIMENTAL RESULTS

#### A. Data

To train and evaluate the performance of the virtual channel EEG signal generative model, the multi-channel dataset from the PRED+CT public database was selected. This database, initially released by Cavanaugh et al. [10] in 2019, includes EEG signals from 119 subjects aged from 18 to 24 years old who completed a probabilistic learning task involving correctly pairing Japanese characters. Anxiety labels were determined based on the Spielberger Trait Anxiety Inventory (TAI), with 60 subjects classified as having no or mild anxiety (TAI < 37) and 59 subjects classified as having moderate or severe anxiety (TAI ⩾ 37). The EEG signals were collected using the Synamps2 system, which has 66 electrodes positioned according to the 10-20 international system, with a sampling frequency of 500 Hz.

Since the frontal, parietal, and temporal lobes are key brain regions associated with emotion regulation and anxiety, particularly the prefrontal cortex [27]-[32], this study selected 17 channels from the dataset, namely Fp1, Fp2, F7, F3, Fz, F4, F8, T3, C3, Cz, C4, T4, T5, P3, Pz, P4, and T6, referred to as the multi-channel dataset for model training.

TABLE II
DATASET INFORMATION

| Dataset | Subject (Male/Female) | Age | Sampling frequency | Channels | Anxiety labels |
|---|---|---|---|---|---|
| Multi-channel dataset | 119 (74 / 45) | 18 ~ 24 | 500 Hz | 17 | No or mild anxiety / Moderate or severe anxiety |

Additionally, using a self-developed portable EEG detector [33], EEG signals were collected from 35 subjects at rest and while wearing a virtual reality (VR) headset watching specific videos. This device can collect EEG signals from four channels (Fp1, Fp2, F7, and F8) with a sampling rate of 250 Hz, and this dataset is called the few-channel dataset. Anxiety states were determined using the Generalized Anxiety Disorder Scale, the Hamilton Anxiety Scale, and the State-Trait Anxiety Inventory, with 21 subjects classified as having no or mild anxiety and 14 subjects classified as having moderate or severe anxiety. The basic information of the two datasets is shown in Table II.

A fourth-order Butterworth bandpass filter was applied to the signals in both datasets to extract EEG signals in the 0.5 ~ 45 Hz frequency band, removing power line and high-frequency interference. Baseline drift was then removed, and the signals were Z-score normalized to accelerate network learning. Finally, the signals were segmented at intervals of 3000 sampling points. The multi-channel dataset was randomly divided into training, validation, and test sets in a 7:1:2 ratio.

The network output included 17 channels, with the spatial distribution of the channels shown in Fig. 2, including the four original channels (solid green circles) and the 13 generated virtual channels (dashed red circles). In addition to the 13 generated virtual channels, the network output also included the four input channels, with constraints strengthened and the model's ability to generate original signals assessed.

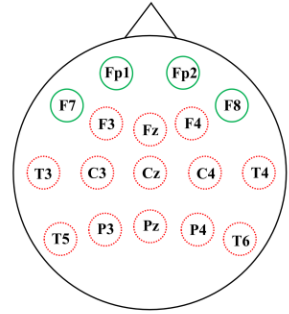

Fig. 2. Spatial position of channels.

The model was trained using the composite weighted loss function in Equation (1), with weight constants α and β both set to 1. The Adam optimizer was used with a learning rate of 0.0001 and momentum parameters of 0.9 and 0.999 to accelerate training and stabilize gradient updates. The batch size was set to 1, and the number of iterations was 11 epochs.

All experiments were completed on a workstation with the following hardware configuration: an Nvidia RTX 4060 GPU, an Intel Core i7-13700H processor, and 32GB of memory. The experiments were conducted in a Python 3.12.4 environment, using the PyCharm 2022 integrated development environment and the TensorFlow (v2.16.1) deep learning framework.

#### B. Experimental Results and Analysis

Quantitative reconstruction metrics across 13 target channels in the multi-channel dataset PRED+CT are summarized in Fig. 3.

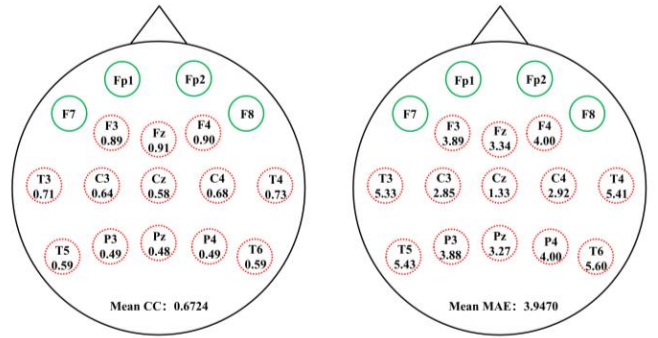

Fig. 3. Performance parameters of the STFPF-EEG GNet.

To verify the effectiveness of the proposed model, reference models highly matching the task were selected for comparison.

The WGAN proposed in [23] only supports single-channel generation, which does not match the multi-channel generation task of this study; the regional division scheme in [25] makes each region's channel signal generation task relatively independent, which does not match the generation task of this study, so they were not included in the comparison. The algorithm in [24] was not included in the direct comparison because the core module's implementation details were not publicly available. However, it was mentioned that its performance was comparable to the model in [22], with slightly better performance in some specific scenarios, which can provide indirect reference. Ultimately, the SR model in [21], the GN2 model in [22], and the Attention U-Net in [26] were chosen as comparison benchmarks due to their high relevance to the structure and task objectives. All models used the same dataset division, and the hyperparameters of each model were independently optimized. Performance evaluation was conducted using the same evaluation metrics. The comparison results of the different models in terms of MAE and are shown in Fig 4.

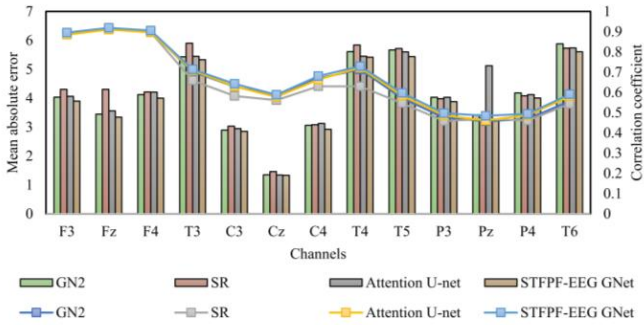

Fig. 4. Performances comparison of different model.

The results indicate that the virtual channel EEG signals generated by the STFPF-EEG GNet have the highest CC values with the corresponding real EEG signals, with an average value of 0.6724. The MAE values between the virtual channel EEG signals and the corresponding real EEG signals are the lowest, with an average of 3.9470, indicating higher signal reconstruction accuracy. Overall, its performance is slightly better than that of the GN2, SR, and attention U-Net models.

Fig. 3 and Fig. 4 show that the spatial distance between the virtual channel and the source channel significantly affects the quality of the generated signal. There is a noticeable spatial decay effect in the correlation between the generated signal and the original signal. The performance parameters of channels F3, Fz, and F4, which are close to the source channels, are optimal, with an average CC value of 0.9. The CC values of channels T3 and T4 exceed 0.7, and the shapes and features of these virtual signals are highly consistent with the original signals. As the spatial distance increases, the CC values of channels T5, P3, P4, and other distant channels drop to around 0.55, indicating a lower consistency between the virtual channel EEG signals and the original signals compared to the near channels. This result is consistent with the findings in [24]. Brain regions that are farther apart have larger functional differences, and the correlation between their EEG signals decreases, making signal generation more challenging. Despite the lower CC values for distant channels, the generated signals still retain certain original signal information components.

Fig. 5 shows a comparison of some time segments of the generated 13 virtual channel EEG signals and the original channel real EEG signals.

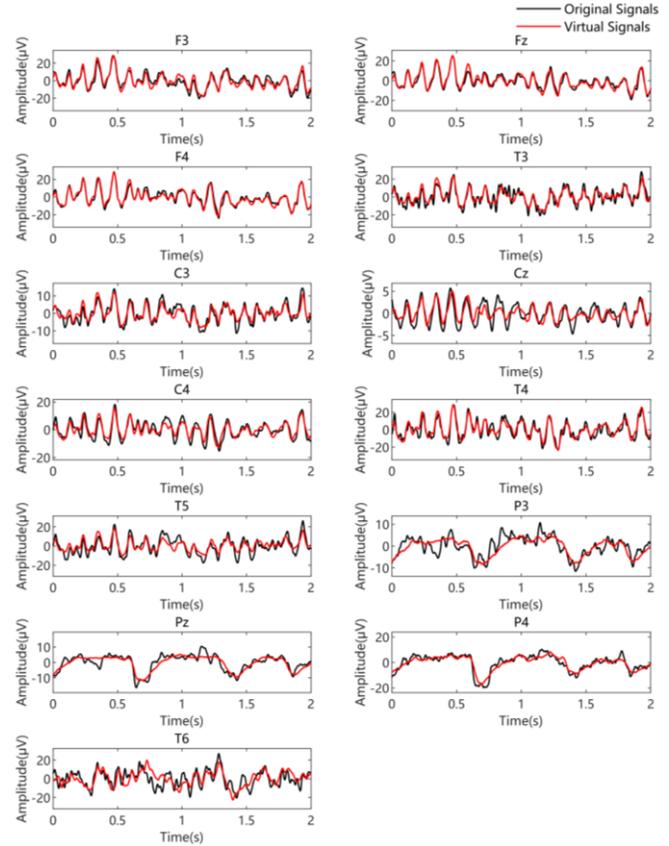

Fig. 5. Comparison of generated virtual EEG signals and original signals.

To further verify the effectiveness of the generated virtual signals, anxiety classification was performed on the multi-channel dataset.

Based on the frequency characteristics of EEG signals [34], [35], bandpass filters with passbands of 8 ~ 14 Hz, 14 ~ 30 Hz, 4 ~ 8 Hz, and 0.5 ~ 4 Hz were used to decompose the signals into alpha, beta, theta, and delta subbands, respectively. Eight types of features were calculated for each subband, including Hjorth parameters (activity, mobility, and complexity), power spectral density ratio, mean absolute amplitude, mean energy, approximate entropy, and fuzzy entropy. Additionally, two features were extracted from the total signal: maximum energy and its corresponding frequency. A total of 34-dimensional features were extracted for each channel signal, and feature dimensionality reduction was performed using analysis of variance to determine the significance of features between different anxiety categories.

The EEG signals were randomly divided into training and test sets in an 8:2 ratio. Three classifiers—K-nearest neighbors(KNN), random forest(RF), and support vector machine(SVM)—were used for binary anxiety state classification, and the classification results with the original channel EEG signals and those with the added virtual channel EEG signals were compared to assess the role of the generated virtual channel EEG signals. Accuracy, precision, recall, and AUC were used to evaluate the performance of each classifier.

In the multi-channel dataset classification, the data were divided into four cases: original 4-channel signals, original 17-channel signals, original 4-channel signals combined with virtual 13-channel signals, and original 60-channel signals. The classification results for the multi-channel dataset are shown in Table III.

TABLE III
ANXIETY CLASSIFICATION RESULTS FOR THE MULTI-CHANNEL DATASET

| Data | Classifiers | Accuracy | Precision | Recall | AUC |
|---|---|---|---|---|---|
| original 4-channel | KNN | 0.7917 | 0.7947 | 0.7555 | 0.7621 |
| | RF | 0.7639 | 0.7952 | 0.7687 | 0.7811 |
| | SVM | 0.8150 | 0.8220 | 0.8044 | 0.8058 |
| original 17-channel | KNN | 0.8889 | 0.9251 | 0.8454 | 0.9111 |
| | RF | 0.8750 | 0.9302 | 0.8888 | 0.9242 |
| | SVM | 0.9028 | 0.9353 | 0.9153 | 0.9586 |
| original 4 + virtual 13 channel | KNN | 0.8611 | 0.8888 | 0.8488 | 0.8811 |
| | RF | 0.8394 | 0.8609 | 0.8222 | 0.8686 |
| | SVM | 0.8838 | 0.9030 | 0.8733 | 0.9062 |
| original 60-channel | KNN | 0.9167 | 0.8444 | 0.9329 | 0.9259 |
| | RF | 0.9206 | 0.8666 | 0.9134 | 0.9568 |
| | SVM | 0.9383 | 0.9285 | 0.9529 | 0.9851 |

As shown in Table III, when 13 virtual channel signals were added, the classification accuracy of the three classifiers was higher than that of the original four-channel signals, with the highest accuracy for anxiety state binary classification reaching 88.38%. The improvement in accuracy, precision, recall, and AUC was approximately 8%, 9%, 8%, and 12%, respectively, indicating that the addition of virtual channel signals significantly increased the accuracy of anxiety state recognition. Compared with the original 17-channel signals, the classification accuracy was slightly lower, especially for the support vector machine, with a difference of 2% in classification results. The precision, recall, and AUC indicators had gaps of less than 6%, indicating that the generated virtual channel signals could provide additional effective information to improve the recognition rate and further approach the performance of the original 60-channel signals.

The anxiety classification results verified the role of the STFPF-EEG GNet model. The generated virtual channel signals effectively supplemented the EEG signal information of the frontal, temporal, and parietal lobes, which are important brain regions, and improved anxiety classification performance.

## IV. CONCLUSIONS

To address the issue of insufficient data due to the limited number of channels in portable EEG devices, this study developed a virtual channel generation method based on spatio-temporal feature fusion. The model mainly includes a temporal domain feature calculation module and a spatial domain feature acquisition module. The temporal domain feature calculation module is used to extract the features of each channel in the time dimension, while the spatial domain feature acquisition module aims to capture the correlation features between multiple channels. The results from experiments using a public anxiety database show that the STFPF-EEG GNet model generated virtual signals with high consistency with the original signals and effectively improved the performance of machine learning models for anxiety recognition. This study provides an effective solution for the application of portable EEG devices in mental health status assessment.

There are still several important issues to be further investigated: 1) In terms of model architecture design, attention mechanisms can be introduced to clarify the attention weights of each convolutional channel for structural simplification and optimization; 2) A more effective spatio-temporal feature fusion strategy may be adopted to replace the current direct stacking fusion method. For example, spatial and temporal attention modules could be used to separately extract the weights of the two types of features, which are then fused and weighted before output, or the weights of different features could be dynamically adjusted to improve the feature fusion effect; 3) For channels farther from the source channels, the generated virtual signals have larger discrepancies with the real signals. Regularization constraints or weights could be added in the channel dimension to strengthen the consistency between the generated signals and the real signals, reducing generation errors.